\documentclass[reqno]{amsart}

\usepackage{amssymb}
\usepackage{amsmath}
\usepackage{algorithm}
\usepackage{algorithmic}
\usepackage{multirow}
\usepackage{hyperref}
\usepackage{graphicx}

\usepackage[pagewise]{lineno}

\begin{document}

\title{An Improved ChaCha Algorithm Based on Quantum Random Number}

\author{Chao Liu$^1$}
\author{Shuai Zhao$^{*1,2,3}$}
\author{ChenHao Jia$^1$}
\author{GengRan Hu$^{1,3}$}
\author{TingTing Cui$^{1,3}$}
\address{$^1$School of Cyberspace, Hangzhou Dianzi University, Hangzhou, 310018, China}
\address{$^2$Pinghu Digital Technology Innovation Institute Co., Ltd., Hangzhou Dianzi University, Jiaxing, 314299, China}
\address{$^3$Zhejiang Provincial Key Laboratory of Sensitive Data Security and Confidentiality Governance, Hangzhou, 310018, China}
\email{zhaoshuai@hdu.edu.cn}

\thanks{This work was supported by the Zhejiang Provincial Natural Science Foundation of China (Grant No. LQ24A050005); the Innovation Program for Quantum Science and Technology (Grant No. 2024ZD0302200).}
\thanks{$^*$Corresponding author: Shuai Zhao}



\date{}

\dedicatory{}

\begin{abstract}

Due to the merits of high efficiency and strong security against timing and side-channel attacks, ChaCha has been widely applied in real-time communication and data streaming scenarios. However, with the rapid development of AI-assisted cryptanalysis and quantum computing technologies, there are serious challenges to the secure implementation of ChaCha cipher. To further strengthen the security of ChaCha cipher, we propose an improved variant based on quantum random numbers, i.e., Quantum Random Number Enhanced ChaCha (QRE-ChaCha). Specifically, the design XORs the initial constants with quantum random numbers and periodically injects quantum random numbers into selected state words during odd rounds to enhance diffusion. Compared with the original ChaCha, the present variant shows stronger resistance to differential attacks and generates a keystream with statistical randomness, thereby offering increased robustness against both classical and quantum attacks. To evaluate the security and performance of the present ChaCha, our analysis proceeds in three main parts. Firstly, we analyze its theoretical security in terms of quantum randomness and attack testing, and conduct differential cryptanalysis with an automated search method based on the Boolean satisfiability problem (SAT). Secondly, we subject the keystream generated by the cipher to randomness tests using the NIST statistical test suite and the GM/T 0005-2021 randomness testing standard. Finally, we assess its encryption and decryption performance by measuring its encryption speed on files of various sizes. According to the results, the present ChaCha is significantly improved to resist differential attacks while maintaining the high efficiency of the original ChaCha cipher, and its keystream successfully passes statistical randomness tests using the NIST and GM/T 0005-2021 standards, meeting cryptographic application requirements.

\end{abstract}

\maketitle

\section{Introduction}\label{sec1}

With the rapid development of information technology, the demand for data security has significantly increased, particularly in modern communication and storage systems. Among the various cryptographic algorithms, ChaCha has been widely adopted in critical protocols, such as Transport Layer Security (TLS)\cite{rfc7905}, due to its exceptional speed and robust security design, securing vast amounts of data in transit and at rest. However, as quantum technologies and AI-assisted cryptanalysis techniques have achieved significant progress, the security of conventional cryptographic schemes, including ChaCha, faces more severe challenges. In response to these challenges, prominence has been obtained in two major research directions: post-quantum cryptography\cite{ref1,ref2,ref3,ref4,ref5}, which focuses on developing new algorithms based on mathematical problems that are conjectured to be intractable for quantum computers, and quantum cryptography\cite{ref6,ref7,ref8,ref9,ref10,ref11,ref12,ref13}, which is rooted in quantum physical principles. Nevertheless, both approaches have significant limitations in enhancing existing symmetric ciphers: Post-quantum cryptography research is largely concerned with asymmetric cryptosystems, while in quantum cryptography, the large-scale deployment of quantum key distribution (QKD) is currently impeded by technological and infrastructural constraints.

Due to the inherent uncertainty in quantum physics, quantum random number possesses intrinsic randomness, which is an important branch of quantum cryptography research, and also one of the quantum technologies with mature applications at present. To enhance the security of classical cryptographic algorithms, it has been highly motivated to combine quantum random numbers with classical cryptographic protocols\cite{ref9}, which on the one hand can extend the application scenarios of quantum random number generators, and on the other hand can improve the security of classical cryptographic protocols. In this work, we draw on a similar idea to enhance the algorithm by applying quantum random numbers to ChaCha stream cipher.

Different from the block cipher, stream ciphers operate on data one bit at a time, enabling faster encryption and decryption as well as inherent parallelism. When the seed key and the pseudorandom numbers exhibit high randomness, the stream cipher can provide strong security guarantees\cite{ref14}. One prominent stream cipher, ChaCha, designed by Daniel J. Bernstein in 2008 as an enhanced variant of Salsa20\cite{ref15}, exemplifies this design philosophy. Although it retains the structural foundation of Salsa20, ChaCha introduces a modified key matrix and a more intricate round function to enhance diffusion and bolster security against cryptanalytic attacks. This refinement, achieved without a significant performance penalty, has cemented ChaCha's status as a robust and efficient cipher, leading to its widespread adoption in real-time communication and data transmission scenarios\cite{rfc7905}. Despite its high performance, widespread adoption, and a robust ARX-based (Addition, Rotation, XOR) design secure against various attacks, numerous security analyses have revealed potential vulnerabilities in the ChaCha cipher\cite{ref50}. As reviewed in Section \ref{subsec1}, existing optimization approaches largely focus on modifying the round function or adjusting internal parameters, but these methods often provide only marginal security improvements against key-recovery attacks, particularly those based on differential cryptanalysis.

To address this issue, as illustrated in Figure \ref{fig1}, we propose an improved ChaCha cipher based on quantum random numbers that significantly improves its security while maintaining the original cipher's high performance. By introducing intrinsic randomness from quantum random numbers into both the seeds and the round function, the proposed scheme increases the randomness and cryptanalytic resistance of the round function, strengthens the security of the generated keystream, and enhances the cipher's resilience against differential cryptanalysis and the resistance of its seeds to quantum attacks. The main contributions of this work are summarized as follows:

\begin{figure}
	\centering
	\includegraphics[width=0.8\linewidth]{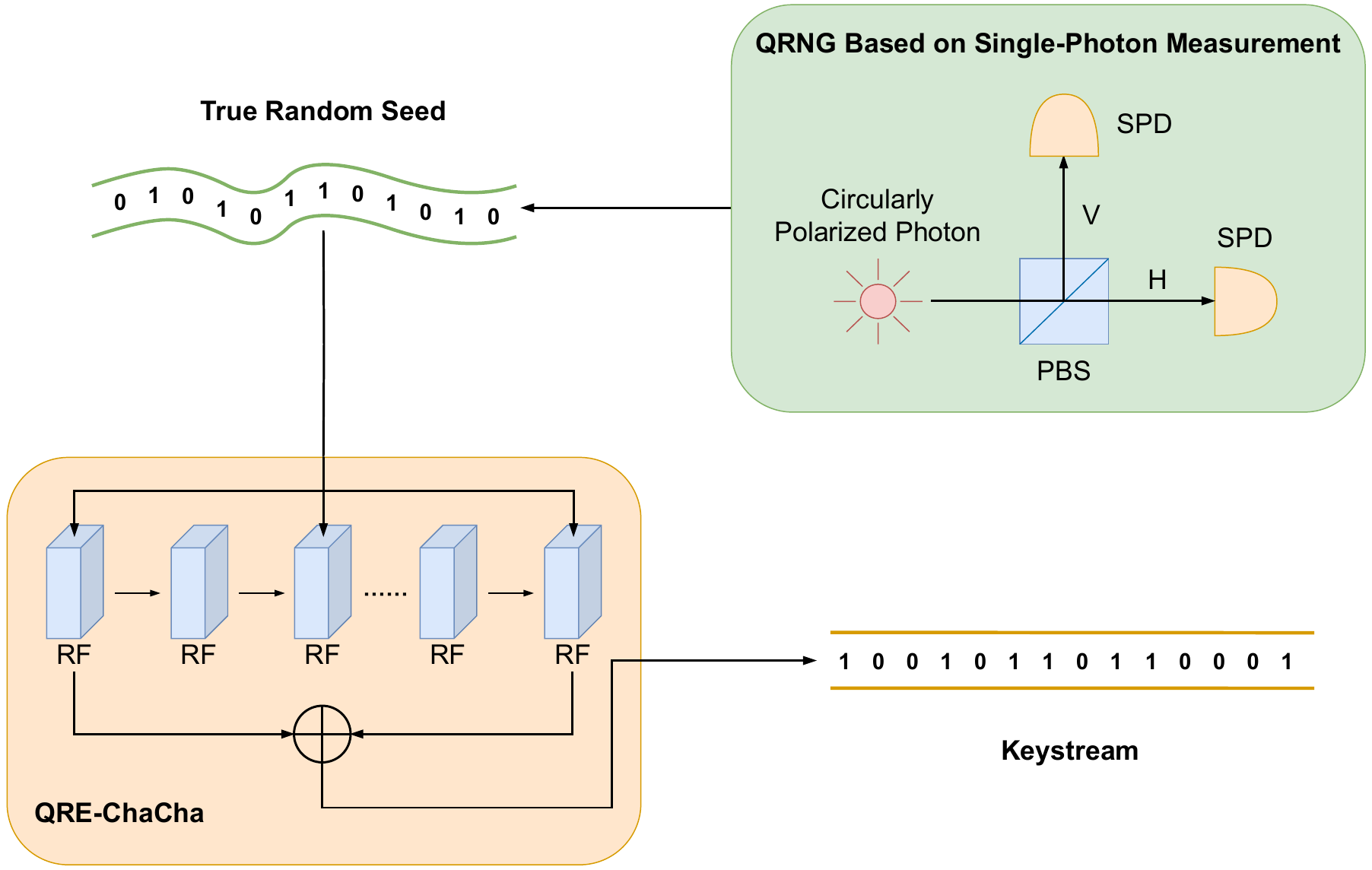}
	\caption{QRE-ChaCha: An integration scheme of quantum random numbers with a classical cipher. \textit{SPD} denotes the single-photon detector, \textit{PBS} denotes the polarization beam splitter, \textit{RF} denotes the round function.}\label{fig1}
\end{figure}

\begin{enumerate}
  \item To address the suboptimal resistance of existing ChaCha optimizations against differential cryptanalysis, we propose a quantum random number enhanced stream cipher, QRE-ChaCha (Quantum Random Number Enhanced ChaCha), which improves the security of the traditional ChaCha cipher by intermittently injecting quantum random numbers into the initial seed and its round function. Specifically, for the initial state matrix, quantum random numbers are XORed with the initial constants; for the intermediate state matrix produced by odd-numbered rounds, quantum random numbers are selectively XORed into specific positions to enhance diffusion. The quantum random numbers are generated through physical processes based on the principles of quantum mechanics, providing true randomness and unpredictability, which significantly strengthens the cipher's resistance to cryptanalytic attacks.
  \item To evaluate the security of the QRE-ChaCha cipher against differential cryptanalysis, we employed an automated search method based on the Boolean satisfiability problem (SAT). According to the results in Figure \ref{fig4}, QRE-ChaCha achieves better differential probabilities for both 2-round and 3-round configurations compared to the original ChaCha cipher, indicating improved resistance to differential attacks. In addition, we conducted statistical randomness tests on the generated keystream using the NIST Statistical Test Suite and the Chinese national standard GM/T 0005-2021 for randomness evaluation. The results show that the keystreams generated by QRE-ChaCha passed both sets of tests, demonstrating compliance with cryptographic standards.
  \item To evaluate the performance of the QRE-ChaCha cipher, we measured its encryption and decryption speeds and compared them with those of the original ChaCha cipher. The purpose of this test was to examine whether the enhanced security features of QRE-ChaCha lead to any significant performance degradation. According to the results, when the time overhead of quantum random number generation is excluded, since the random numbers can be pre-stored in the memory, as shown in Figure \ref{fig3}, the encryption and decryption speeds of QRE-ChaCha are nearly identical to those of ChaCha.
\end{enumerate}

The remainder of this paper is organized as follows: In Section \ref{sec2}, we provide a brief overview of related work and recent advances in the study of ChaCha and quantum random numbers. To support the proposed design, we introduce the necessary preliminaries in Section \ref{sec3}. In Section \ref{sec4}, we detail the QRE-ChaCha enhancement scheme. In Section \ref{sec5}, to analyze the security of QRE-ChaCha, we describe its theoretical properties and present the results of our differential cryptanalysis. In Section \ref{sec6}, we present the methodology and results of randomness testing. A performance evaluation is then provided in Section \ref{sec7}. Finally, we conclude this work in Section \ref{sec8}.

\section{Related Work}\label{sec2}

\subsection{Quantum Random Number Generation}

In the field of quantum random number generation, ensuring the security and reliability of randomness sources has attracted a lot of research interest. As summarized by Mannalatha et al.\cite{ref16} and Herrero-Collantes et al.\cite{ref18}, quantum random number generators (QRNGs) offer intrinsic unpredictability and true randomness based on quantum physics, which are fundamentally inexplicable by classical theories. Recent advances have led to the development of device-independent (DI) and semi-device-independent (SDI) QRNG models, which aim to minimize or eliminate trust assumptions on the quantum devices. These models strengthen resistance against side-channel attacks by ensuring that the randomness generation process cannot be tampered with or predicted even when parts of the device are untrusted. Ma et al.\cite{ref17} further classified QRNGs into practical, SDI, and DI types, and analyzed their trade-offs in terms of implementation complexity, generation rate, and provable security. Overall, DI and SDI QRNGs have become key approaches for constructing secure randomness sources under different security assumptions, particularly suitable for enhancing cryptographic primitives that demand high levels of entropy and adversarial robustness.

In terms of application domains for quantum random numbers, Iavich et al.\cite{ref19} proposed a novel QRNG in 2020 based on photon arrival time, aiming to generate fast and high-quality quantum random numbers at a low cost to meet the demands of cryptographic applications. To combine the inherent randomness of quantum processes with the efficiency of classical pseudorandom number generation, they further introduced a hybrid QRNG\cite{ref20} in 2021, integrating techniques from time-of-arrival QRNGs, photon-counting QRNGs, and attenuated-pulse QRNGs. It was specifically designed for use in cryptographic algorithms. Experimental results demonstrated its ability to produce high-quality random numbers at a relatively high throughput, making it highly effective in conventional cryptographic applications. Similarly, Stipcevic et al.\cite{ref21} focused explicitly on the cryptographic context, emphasizing the significance of random number generators in secure systems. By comparing free-running oscillator-based RNGs with QRNGs, their work explored the role of randomness in quantum cryptography, discussed various post-processing techniques, and proposed evaluation methods, thereby offering guidance on practical QRNG deployment in cryptographic settings.

Distinct from hardware-based QRNGs, Kuang et al.\cite{ref22} proposed a pseudo-quantum random number generator (pQRNG) based on quantum algorithms. Leveraging the high-entropy properties of quantum permutation spaces, the proposed method uses Quantum Permutation Pad (QPP) techniques to generate pseudorandom numbers with strong unpredictability. Without the need for physical quantum integration, the pQRNG can be readily embedded into classical computing systems, providing a high-quality deterministic randomness source for cryptographic and other applications.

To support real-world adoption of quantum random numbers, Huang et al.\cite{ref23} developed a practical cloud-based quantum random number service by integrating QRNGs with Alibaba Cloud infrastructure. This platform provides quantum-grade randomness for a variety of applications, including cryptographic systems.

It is worth noting that in 2023, JianWei Pan et al.\cite{ref9} proposed an enhanced zero-knowledge proof scheme based on device-independent quantum randomness, further demonstrating the feasibility of integrating quantum random numbers with classical cryptographic protocols. In their work, a quantum solution was introduced in the form of a quantum randomness service, which generates random numbers via loophole-free Bell tests and transmits them using post-quantum cryptographic authentication, thereby improving the overall security of the protocol. Inspired by similar principles, the present study introduces quantum random numbers into the round function of the ChaCha cipher to enhance its security, leading to the design of the QRE-ChaCha.

\subsection{ChaCha cipher}\label{subsec1}

As a member of the stream cipher family, the ChaCha has been widely adopted due to its high performance and simplicity. To date, most studies have focused on structural analysis and cryptanalytic attacks against the cipher\cite{ref24,ref25,ref26,ref27,ref28,ref29,ref30,ref31,ref32,ref33,ref34,ref35}, while relatively fewer works have explored the optimization of ChaCha's design and performance\cite{ref36,ref37,ref38,ref39}.

With regard to the analysis of the ChaCha, Najm et al.\cite{ref24} compared AES and ChaCha20 on microcontrollers in order to assess the side-channel resistance of ChaCha. Their results showed that, although ChaCha20 exhibits stronger resistance to side-channel attacks than AES, it incurs higher computational overhead. To reveal potential weaknesses of ChaCha20 in terms of fault resistance, Kumar et al.\cite{ref25} were the first to introduce a practical fault attack against the ChaCha20 stream cipher and proposed four differential fault analysis techniques. To lay a theoretical foundation for the practical deployment of ChaCha20-Poly1305, Procter et al.\cite{ref14} analyzed the security of combining ChaCha20 with the Poly1305 authenticator in IETF protocols and provided a security reduction for the authenticated encryption scheme. To further analyze the security of ChaCha20-Poly1305 in multi-user environments, Degabriele et al.\cite{ref26} proposed an enhanced analytical framework, established tighter security bounds and adversarial models, and identified security limits of the scheme under such settings. To study the rotational properties of the ChaCha permutation, Barbero et al.\cite{ref27} proposed a distinguishing method based on permutation calls and derived upper and lower bounds on the probability of rotation propagation. However, the authors noted that no effective cryptanalytic applications of these results to the ChaCha stream cipher have yet been found. To compare AES, 3DES, and ChaCha20 in terms of performance and efficiency, Claros et al.\cite{ref28} designed an experiment and found that ChaCha20 achieved the fastest encryption and decryption speeds.

Innovative analytical approaches to the ChaCha can be traced back to the introduction of the probabilistic neutral bits (PNB) concept by Aumasson et al.\cite{ref29} in 2008. This concept was first applied to differential cryptanalysis of the Salsa20 and ChaCha stream ciphers, leading to successful attacks and marking the first time this technique was used in cryptanalysis. Subsequently, to reduce the time and data complexity of attacks against reduced-round variants of Salsa20 and ChaCha, Shi et al.\cite{ref30} proposed an improved key-recovery method by introducing new types of distinguishers and identifying high-probability second-order differential paths. To evaluate the differential security of Salsa and ChaCha using a mixed analytical framework, Choudhuri et al.\cite{ref31} developed a hybrid model that applies original nonlinear functions in the initial rounds and linear approximations in the later rounds, concluding that 12 rounds suffice to protect a 256-bit key. To correct flaws in prior analyses and extend attacks to other reduced-round versions, Deepthi et al.\cite{ref32} conducted a thorough reanalysis of ChaCha20 and Salsa20 and successfully mounted attacks on Salsa20/7, ChaCha6, and ChaCha7. To enhance the effectiveness of PNB-based differential attacks, Miyashita et al.\cite{ref33} introduced an optimized PNB differential attack, identifying the best differential biases and combinations of neutral bits for ChaCha. Later, to improve the precision of differential-linear cryptanalysis, Bellini et al.\cite{ref34} expanded the search space, optimized mask selection between differential and linear components, and utilized MILP tools to construct distinguishers against 7, 7.5, and 5 rounds of ChaCha. To explore the potential of higher-order differential-linear attacks on ChaCha, Ghafoori et al.\cite{ref35} analyzed multiple round configurations and discovered new differential biases and linear approximations, further advancing the complexity and scope of ChaCha cryptanalysis.

In the domain of ChaCha optimization, several efforts have been made to enhance its security and adaptability across different application scenarios. To improve the security of ChaCha while ensuring low power consumption for IoT devices, Mahdi et al.\cite{ref36} proposed an enhanced variant called Super ChaCha, which modifies the rotation process and the input update order to increase resistance against cryptanalytic attacks. Similarly, to enhance data privacy on the Internet of Things platforms, Jain et al.\cite{ref37} presented an optimized version of ChaCha20 tailored for secure communication in constrained environments. To strengthen the security of ChaCha20 through structural improvements, Kebande et al.\cite{ref38} introduced EChaCha20, which enhances the quarter-round function (QR-F) and incorporates 32-bit input words along with ARX operations. To apply ChaCha in lightweight multimedia encryption, Maolood et al.\cite{ref39} proposed a novel video encryption scheme that integrates the ChaCha20 stream cipher with hybrid chaotic mapping theory, achieving efficiency and lightweight security suitable for real-time video processing.

\section{Preliminaries}\label{sec3}

To facilitate subsequent reading and understanding, we first present the relevant notational conventions, as summarized in Table \ref{tab1}. Then, a brief overview of the ChaCha cipher is provided.

\begin{table}
    \centering
    \caption{Notation Conventions.}\label{tab1}
    \begin{tabular}{|c|c|}
        \hline
        \textbf{Symbol} & \textbf{Definition} \\
        \hline
        $X$ & A $4\times4$ ChaCha matrix composed of 16 words \\
        $X^{(0)}$ & Initial state matrix of ChaCha/QRE-ChaCha \\
        $X^{^{\prime}(0)}$ & Related matrix with a one-bit difference at position $x_{i,j}$ \\
        $X^{(R)}$ & ChaCha/QRE-ChaCha matrix after $R$ rounds \\
        $X^{(r)}$ & ChaCha/QRE-ChaCha matrix after $r$ rounds, where $R > r$ \\
        $x_i^{(R)}$ & The $i$-th word of the state matrix $X^{(R)}$ \\
        $x_{i,j}^{(R)}$ & The $j$-th bit of the $i$-th word in $X^{(R)}$ \\
        $p$ & Probability of a differential path \\
        $x \boxplus y$ & Modular addition of words $x$ and $y$ \\
        $x \boxminus y$ & Modular subtraction of words $x$ and $y$ \\
        $x\oplus y$ & Bitwise XOR of words $x$ and $y$ \\
        $x\lll n$ & Left rotation of word $x$ by $n$ bits \\
        $\Delta x$ & XOR difference between word $x$ and $x^{\prime}$ \\
        $\text{ChaCha n}$ & ChaCha stream cipher at round $n$ \\
        $\text{QRE-ChaCha n}$ & QRE-ChaCha stream cipher at round $n$ \\
        \hline
    \end{tabular}
\end{table}

ChaCha is a pseudorandom function based on the ARX structure. It updates its internal state matrix by performing four modular additions, four XOR operations, and four bitwise rotations. Compared to Salsa, ChaCha updates each word twice per round rather than once, which enhances diffusion across the state matrix and improves resistance to cryptanalysis. ChaCha follows the same design principles as Salsa, using 32 bits (one word) to construct a 512-bit (16-word) initial state matrix, which consists of four constant words ($c_1=0x61707865$, $c_2=0x3320646e$, $c_3=0x79622d32$, $c_4=0x6b206574$), eight seed key words, three nonce words, and one counter word.

As an iterative stream cipher, the number of rounds in ChaCha can be selected based on the desired level of security and performance. For maximum security, 20 rounds are used; for maximum speed, 8 rounds may be chosen; and for a trade-off between the two, 12 rounds are commonly applied\cite{ref15}. In each round, the ChaCha round function ($R$) is composed of four quarter-round functions ($QR$), each of which takes four input words $(x_a^{(r)}, x_b^{(r)}, x_c^{(r)}, x_d^{(r)})$.

The initial state matrix of ChaCha is shown in Equation (\ref{eq1}):

\begin{equation}\label{eq1}
    X^{(0)} =
    \begin{pmatrix}
        x_{0}^{(0)}  & x_{1}^{(0)}  & x_{2}^{(0)}  & x_{3}^{(0)}  \\
        x_{4}^{(0)}  & x_{5}^{(0)}  & x_{6}^{(0)}  & x_{7}^{(0)}  \\
        x_{8}^{(0)}  & x_{9}^{(0)}  & x_{10}^{(0)} & x_{11}^{(0)} \\
        x_{12}^{(0)} & x_{13}^{(0)} & x_{14}^{(0)} & x_{15}^{(0)}
    \end{pmatrix}
    =
    \begin{pmatrix}
        c_{0} & c_{1} & c_{2} & c_{3} \\
        k_{0} & k_{1} & k_{2} & k_{3} \\
        k_{4} & k_{5} & k_{6} & k_{7} \\
        t_{0} & \nu_{0} & \nu_{1} & \nu_{2}
    \end{pmatrix},
\end{equation}

In Equation (\ref{eq1}), $c$ denotes the constant, $k$ denotes the seed key, $t$ denotes the counter, and $v$ denotes the nonce.

\begin{figure}
	\centering
	\includegraphics[width=0.5\linewidth]{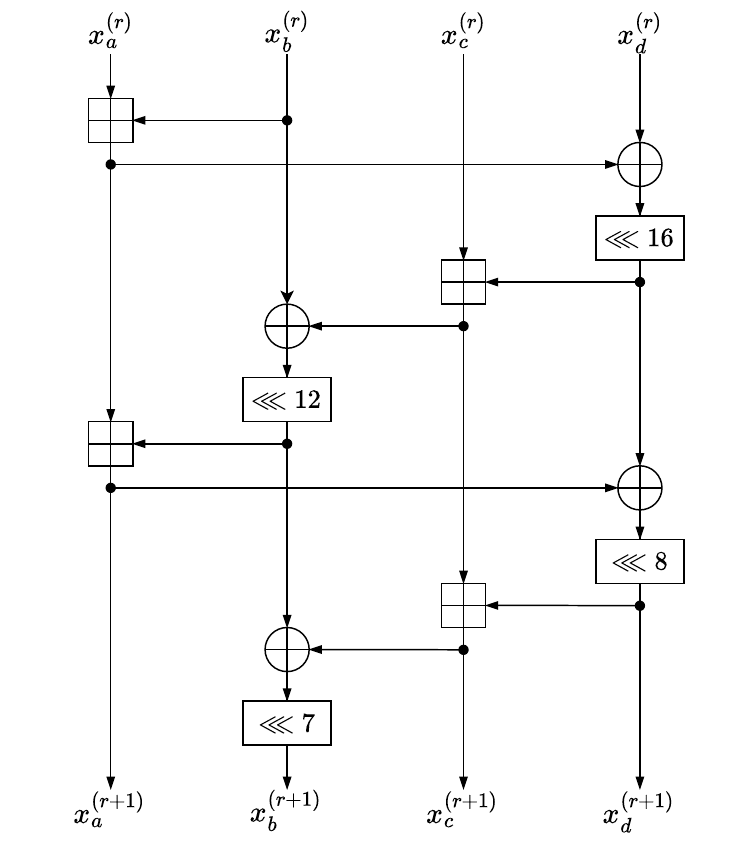}
	\caption{The quarter-round function structure of the ChaCha cipher.}\label{fig2}
\end{figure}

The overall structure of the quarter-round function is illustrated in Figure \ref{fig2}. Since the four input words of each quarter-round function in a round are distinct, ChaCha's round function allows all four quarter-rounds to be executed in parallel. For odd-numbered rounds (column rounds), the quarter-round functions operate on the following four column vectors respectively: $(x_0^{(r)}, x_4^{(r)}, x_8^{(r)}, x_{12}^{(r)})$, $(x_1^{(r)}, x_5^{(r)}, x_9^{(r)}, x_{13}^{(r)})$, $(x_2^{(r)}, x_6^{(r)}, x_{10}^{(r)}, x_{14}^{(r)})$, $(x_3^{(r)}, x_7^{(r)}, x_{11}^{(r)}, x_{15}^{(r)})$. For even-numbered rounds (diagonal rounds), the quarter-round functions are applied to the following four diagonal vectors respectively: $(x_0^{(r)}, x_5^{(r)}, x_{10}^{(r)}, x_{15}^{(r)})$, $(x_1^{(r)}, x_6^{(r)}, x_{11}^{(r)}, x_{12}^{(r)})$, $(x_2^{(r)}, x_7^{(r)}, x_8^{(r)}, x_{13}^{(r)})$, $(x_3^{(r)}, x_4^{(r)}, x_9^{(r)}, x_{14}^{(r)})$.

Each quarter-round updates the internal state matrix $X^{(r)}$ by sequentially applying the operations defined in Equation (\ref{eq2}) to its four input words.

\begin{equation}\label{eq2}
    \begin{cases}
        \begin{aligned}
            x_{a'}^{(r)} &= x_a^{(r)} \boxplus x_b^{(r)} &
            x_{d'}^{(r)} &= x_d^{(r)}\oplus x_{a'}^{(r)} &
            x_{d''}^{(r)} &= x_{d'}^{(r)}\lll16 \\
            x_{c^\prime}^{(r)} &= x_c^{(r)} \boxplus x_{d^{\prime\prime}}^{(r)} &
            x_{b^{\prime}}^{(r)} &= x_b^{(r)}\oplus x_{c^{\prime}}^{(r)} &
            x_{b^{\prime\prime}}^{(r)}&=x_{b^{\prime}}^{(r)}\lll12 \\
            x_a^{(r+1)} &= x_{a^\prime}^{(r)} \boxplus x_{b^{\prime\prime}}^{(r)} &
            x_{d^{\prime\prime\prime}}^{(r)} &= x_{d^{\prime\prime}}^{(r)}\oplus x_a^{(r+1)} &
            x_d^{(r+1)} &= x_{d^{\prime\prime\prime}}^{(r)}\lll8 \\
            x_c^{(r+1)} &= x_{c^{\prime}}^{(r)} \boxplus x_d^{(r+1)} &
            x_{b^{\prime\prime\prime}}^{(r)} &= x_{b^{\prime\prime}}^{(r)}\oplus x_c^{(r+1)} &
            x_b^{(r+1)}&=x_{b^{\prime\prime}}^{(r)}\lll7
        \end{aligned}
    \end{cases}.
\end{equation}
\vspace{2ex}

For the $n$-round version of ChaCha, the final 512-bit pseudorandom keystream block $Z$ is computed as: $$Z = X^{(0)} + X^{(n)}.$$

\section{Quantum Random Number Enhanced ChaCha}\label{sec4}

In order to strengthen the security of the original ChaCha cipher by introducing true randomness, the proposed Quantum Random Number Enhanced ChaCha (QRE-ChaCha) modifies the round transformation mechanism. Specifically, quantum random numbers are first XORed with the constant words in the initial state matrix. Then, after each odd-numbered round, additional quantum random numbers are XORed into the first 128 bits of the intermediate state matrix.

\begin{figure}
	\centering
	\includegraphics[width=1\linewidth]{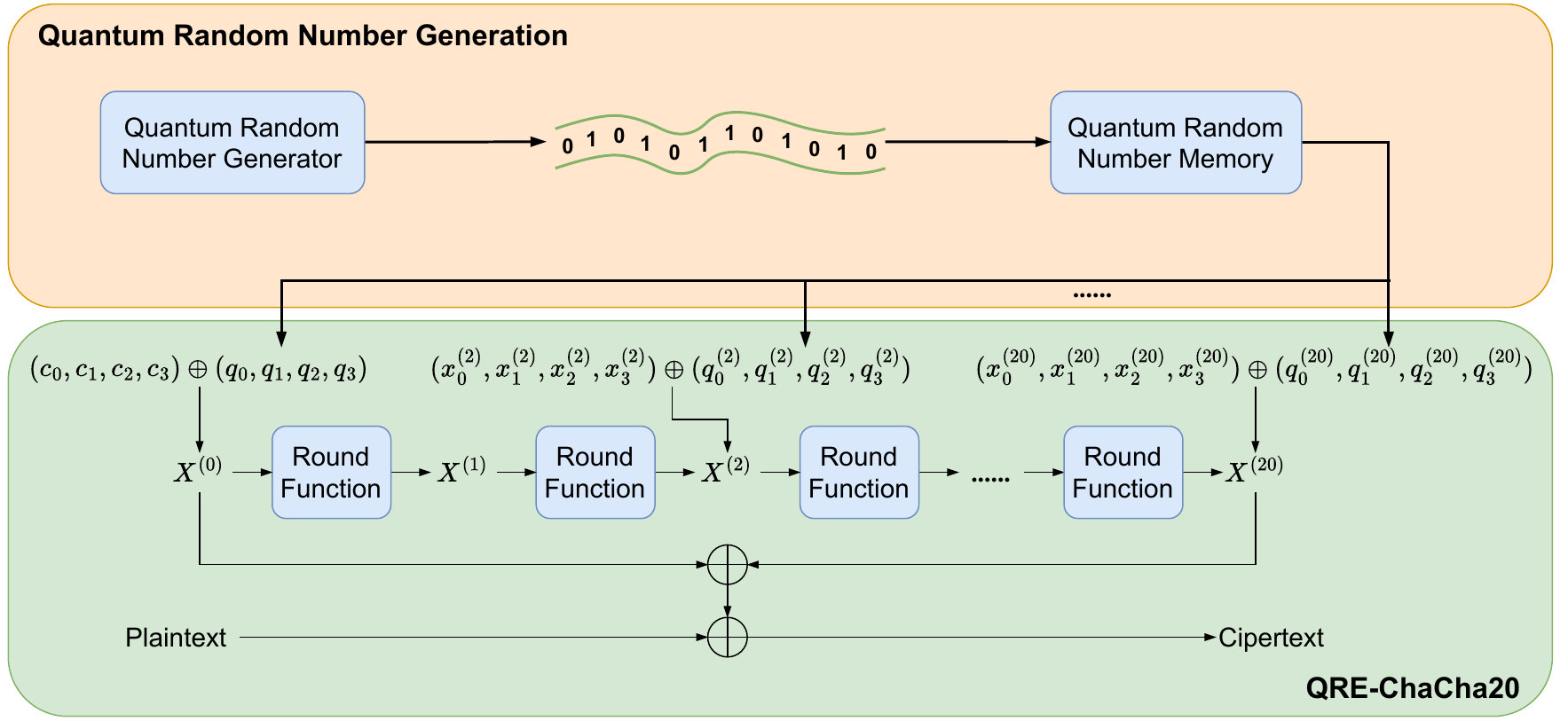}
	\caption{The overall process of the 20-round QRE-ChaCha cipher.}\label{fig3}
\end{figure}

As illustrated in Figure \ref{fig3}, the quantum random number generator produces truly random bits and stores them in a quantum random number memory module, which is subsequently accessed by the QRE-ChaCha cipher. It is worth noting that, since this work focuses solely on the optimization of the ChaCha cipher, the quantum random number generation process itself is not illustrated in detail in Figure \ref{fig3}, as it merely serves as a service module invoked by the algorithm and can be pre-stored in the memory module.

The detailed optimization strategy of QRE-ChaCha is summarized as follows:

\begin{itemize}
    \item For the initial state matrix, QRE-ChaCha replaces the 128-bit (4-word) constant $(c_0,c_1,c_2,c_3)$ with the bitwise XOR of this constant and a 128-bit quantum random number $q$, while the remaining parts use a 256-bit (8-word) secret key, a 96-bit (3-word) nonce, and a 32-bit  counter as input, as shown in Equation (\ref{eq3}), where $q$ denotes the quantum random number.
    \item For each odd-numbered round of the quarter-round (QR) function, QRE-ChaCha modifies the input state matrix (i.e., the output state matrix of the preceding even-numbered round) by XORing the first four words $(x_0^{(r-1)},x_1^{(r-1)},x_2^{(r-1)},x_3^{(r-1)})$ with four quantum random words of the same length $(q_0^{(r-1)},q_1^{(r-1)},q_2^{(r-1)},q_3^{(r-1)})$, and then replacing the original words with the resulting values. This operation enhances the diffusion and randomness of the round function, as shown in Equation (\ref{eq4}). Here, $r$ starts from 0, so $X^{(r)}$ refers to the input state matrix of an odd-numbered round when $r$ is even. The complete QRE-ChaCha algorithm is described in Algorithm \ref{alg1}.
\end{itemize}

\begin{align}\label{eq3}
    X^{(0)} &=
    \begin{pmatrix}
        x_{0}^{(0)}  & x_{1}^{(0)}  & x_{2}^{(0)}  & x_{3}^{(0)}  \\
        x_{4}^{(0)}  & x_{5}^{(0)}  & x_{6}^{(0)}  & x_{7}^{(0)}  \\
        x_{8}^{(0)}  & x_{9}^{(0)}  & x_{10}^{(0)} & x_{11}^{(0)} \\
        x_{12}^{(0)} & x_{13}^{(0)} & x_{14}^{(0)} & x_{15}^{(0)}
    \end{pmatrix}
    \nonumber \\ &=
    \begin{pmatrix}
        c_{0} \oplus q_{0}^{(0)} & c_{1} \oplus q_{1}^{(0)} & c_{2} \oplus q_{2}^{(0)} & c_{3} \oplus q_{3}^{(0)} \\
        k_{0} & k_{1} & k_{2} & k_{3} \\
        k_{4} & k_{5} & k_{6}  & k_{7} \\
        t_{0} & \nu_{0} & \nu_{1} & \nu_{2}
    \end{pmatrix}.
\end{align}

\begin{align}\label{eq4}
    X^{(r=\text{even})} &=
    \begin{pmatrix}
        x_{0}^{(r)}  & x_{1}^{(r)}  & x_{2}^{(r)}  & x_{3}^{(r)}  \\
        x_{4}^{(r)}  & x_{5}^{(r)}  & x_{6}^{(r)}  & x_{7}^{(r)}  \\
        x_{8}^{(r)}  & x_{9}^{(r)}  & x_{10}^{(r)} & x_{11}^{(r)} \\
        x_{12}^{(r)} & x_{13}^{(r)} & x_{14}^{(r)} & x_{15}^{(r)}
    \end{pmatrix}
    \nonumber \\ &=
    \begin{pmatrix}
        x_{0}^{(r)} \oplus q_{0}^{(r)} & x_{1}^{(r)} \oplus q_{1}^{(r)} & x_{2}^{(r)} \oplus q_{2}^{(r)} & x_{3}^{(r)} \oplus q_{3}^{(r)} \\
        x_{4}^{(r)} & x_{5}^{(r)} & x_{6}^{(r)} & x_{7}^{(r)} \\
        x_{8}^{(r)} & x_{9}^{(r)} & x_{10}^{(r)} & x_{11}^{(r)} \\
        x_{12}^{(r)} & x_{13}^{(r)} & x_{14}^{(r)} & x_{15}^{(r)}
    \end{pmatrix}.
\end{align}

\begin{algorithm}[htb]
    \renewcommand{\algorithmicrequire}{\textbf{Input:}}
    \renewcommand{\algorithmicensure}{\textbf{Output:}}
    \begin{algorithmic}[1]
        \REQUIRE Input parameters Matrix $X$, rounds $R$, QRN $Q$
        \ENSURE Output Keystream $Z$
        \FOR{$r = 0$ \TO $R-1$}
            \IF{$r$ is odd}
                \STATE $(x_0^{(r+1)},x_4^{(r+1)},x_8^{(r+1)},x_{12}^{(r+1)})\gets QR(x_0^{(r)},x_4^{(r)},x_8^{(r)},x_{12}^{(r)})$
                \STATE $(x_1^{(r+1)},x_5^{(r+1)},x_9^{(r+1)},x_{13}^{(r+1)})\gets QR(x_1^{(r)},x_5^{(r)},x_9^{(r)},x_{13}^{(r)})$
                \STATE $(x_2^{(r+1)},x_6^{(r+1)},x_{10}^{(r+1)},x_{14}^{(r+1)})\gets QR(x_2^{(r)},x_6^{(r)},x_{10}^{(r)},x_{14}^{(r)})$
                \STATE $(x_3^{(r+1)},x_7^{(r+1)},x_{11}^{(r+1)},x_{15}^{(r+1)})\gets QR(x_3^{(r)},x_7^{(r)},x_{11}^{(r)},x_{15}^{(r)})$
            \ENDIF
            \IF{$r$ is even}
                \STATE $(x_0^{(r)},x_1^{(r)},x_2^{(r)},x_3^{(r)})\gets (x_0^{(r)},x_1^{(r)},x_2^{(r)},x_3^{(r)})\oplus(q_0^{(r)},q_1^{(r)},q_2^{(r)},q_3^{(r)})$
                \STATE $(x_0^{(r+1)},x_5^{(r+1)},x_{10}^{(r+1)},x_{15}^{(r+1)})\gets QR(x_0^{(r)},x_5^{(r)},x_{10}^{(r)},x_{15}^{(r)})$
                \STATE $(x_1^{(r+1)},x_6^{(r+1)},x_{11}^{(r+1)},x_{12}^{(r+1)})\gets QR(x_1^{(r)},x_6^{(r)},x_{11}^{(r)},x_{12}^{(r)})$
                \STATE $(x_2^{(r+1)},x_7^{(r+1)},x_8^{(r+1)},x_{13}^{(r+1)})\gets QR(x_2^{(r)},x_7^{(r)},x_8^{(r)},x_{13}^{(r)})$
                \STATE $(x_3^{(r+1)},x_4^{(r+1)},x_9^{(r+1)},x_{14}^{(r+1)})\gets QR(x_3^{(r)},x_4^{(r)},x_9^{(r)},x_{14}^{(r)})$
            \ENDIF
        \ENDFOR
        \RETURN $Z = X^{(0)} + X^{(R)}$
    \end{algorithmic}
    \caption{QRE-ChaCha}\label{alg1}
\end{algorithm}

In addition, to ensure the injected quantum randomness effectively propagates across the entire state matrix, we define that the difference of the quantum random numbers is not equal to the difference of the corresponding input word to the quarter-round function, as follows: $$\Delta q_{i}^{(r)} \ne \Delta x_a^{(r)},$$
where $i = 0, 1, 2, 3$, $r$ denotes the $r$-th round of the QRE-ChaCha cipher, and $x_a^{(r)}$ represents the first input word of the quarter-round function. This constraint is an integral part of the algorithm design.

Without loss of generality, it is important to note that the quantum random numbers used in the cipher are assumed to be confidential to adversaries. The distribution method of these quantum random numbers is not specified in this work. In practice, to ensure the security of quantum random number transmission, a recommended approach is to adopt the method used in \cite{ref9}, where the generated random numbers are authenticated using a post-quantum cryptographic (PQC) signature scheme. This approach provides a certain level of resistance against quantum attacks during the distribution process.

\section{Security Analysis}\label{sec5}

\subsection{Quantum Randomness Analysis}

The design of QRE-ChaCha is inspired by the non-interactive zero-knowledge proof (NIZKP) protocol based on device-independent quantum randomness, as proposed in \cite{ref9}. In this work, the security of classical cryptographic schemes is improved by leveraging the intrinsic unpredictability of quantum-generated randomness. Similarly, we incorporate quantum random numbers into the ChaCha cipher to strengthen its cryptographic security in our work.

Quantum random numbers are generated based on fundamental principles of quantum mechanics, which are inherently unpredictable and truly random. In contrast, conventional cryptographic systems typically rely on pseudorandom number generators (PRNGs). Although PRNGs can produce uniformly distributed outputs, as pointed out in \cite{ref18}, uniformity alone is far from sufficient in modern cryptographic applications. Random numbers are now expected to satisfy at least two additional properties: unpredictability (forward security) and backward security. Since PRNGs are inherently deterministic, they cannot provide true randomness and thus may fall short in satisfying these essential security requirements.

For quantum random numbers, their true randomness originates from quantum processes that disrupt coherent superposition states\cite{ref17}. In the most widely used practical QRNGs based on photonic systems, a single photon can carry one quantum bit (qubit), which may be viewed as a linear superposition of the classical bit values 0 and 1, expressed as $(|0\rangle + |1\rangle)/\sqrt{2}$. Upon measurement, the qubit collapses into either 0 or 1 with equal probability (50\%), thus producing a genuinely random binary outcome. In practical implementations, as the example shown in the QRNG part of Figure \ref{fig1}, the photon is initially prepared in a superposition of horizontal (H) and vertical (V) polarization states, denoted as $(|H\rangle + |V\rangle)/\sqrt{2}$. A polarization beam splitter (PBS) is used to transmit horizontally polarized photons and reflect vertically polarized ones. Two single-photon detectors (SPDs), positioned at the output ports of the PBS, are used to measure the outcome. This configuration enables the generation of random bits with theoretically perfect randomness, where the unpredictability is fundamentally guaranteed by the laws of quantum physics.

In terms of attack testing against random number generators, as summarized in \cite{ref40}, QRNGs produce truly random and non-reproducible values, making it impossible to predict future outputs based on past values. In contrast, PRNGs generate sequences based on an initial seed; if the seed is weak or compromised, the entire sequence becomes predictable and vulnerable to attack. QRNGs, in contrast, use quantum entropy sources, where any attempt to probe or tamper with the system inherently disturbs the quantum state, thus making such attacks detectable. Therefore, the true randomness provided by QRNGs significantly enhances the strength of cryptographic keys, reduces the effectiveness of statistical attacks and cryptanalysis that exploit key generation patterns, and adds an extra layer of security to cryptographic systems. Additionally, as shown in \cite{ref41}, QRNGs demonstrate substantial advantages in randomness, uniformity, and resistance to correlation-based attacks through comprehensive statistical evaluations, including NIST, Diehard, and ENT test suites, particularly in scenarios requiring high bit-rate random number generation.

In summary, by integrating quantum random numbers into both the seed initialization and the round function, QRE-ChaCha leverages the intrinsic unpredictability and high entropy of quantum randomness. This design not only improves the statistical quality of the generated keystreams, but also enhances the cipher's robustness against differential cryptanalysis and potential quantum adversaries, thereby reinforcing its theoretical cryptographic security.

\subsection{Differential Cryptanalysis}

In this section, to demonstrate the enhanced security of the proposed QRE-ChaCha against differential attacks, we conduct a detailed cryptanalysis of its reduced-round versions and present the corresponding analysis process and experimental results. Despite the wide range of available cryptanalytic techniques, differential analysis remains one of the most widely adopted methods in symmetric cipher evaluation. Due to practical limitations such as computational resources and device constraints, our differential analysis focuses on the 2-round and 3-round versions of QRE-ChaCha. Nevertheless, the results are sufficient to demonstrate that QRE-ChaCha exhibits improved resistance to differential attacks.

To evaluate the resistance of the QRE-ChaCha against differential cryptanalysis, we adopt the theoretical framework proposed by Kai Fu et al.(2016), which models the differential characteristics and linear approximations of modular addition operations in ARX ciphers using linear inequalities under the assumptions of independently distributed inputs and independent rounds\cite{ref42}. Based on this framework, we utilize the Boolean Satisfiability Problem (SAT)-based automated search method to explore optimal differential characteristics for the reduced-round versions of QRE-ChaCha.The specific differential model and testing tool used are based on the open-source CryptoSMT project developed by Stefan K{\"o}lbl\cite{ref43}. In addition, we used the quantum random number API provided by ETH Z{\"U}RICH\cite{ref44} to obtain the quantum random numbers required for the experiments.

\begin{figure}
	\centering
	\includegraphics[width=0.8\linewidth]{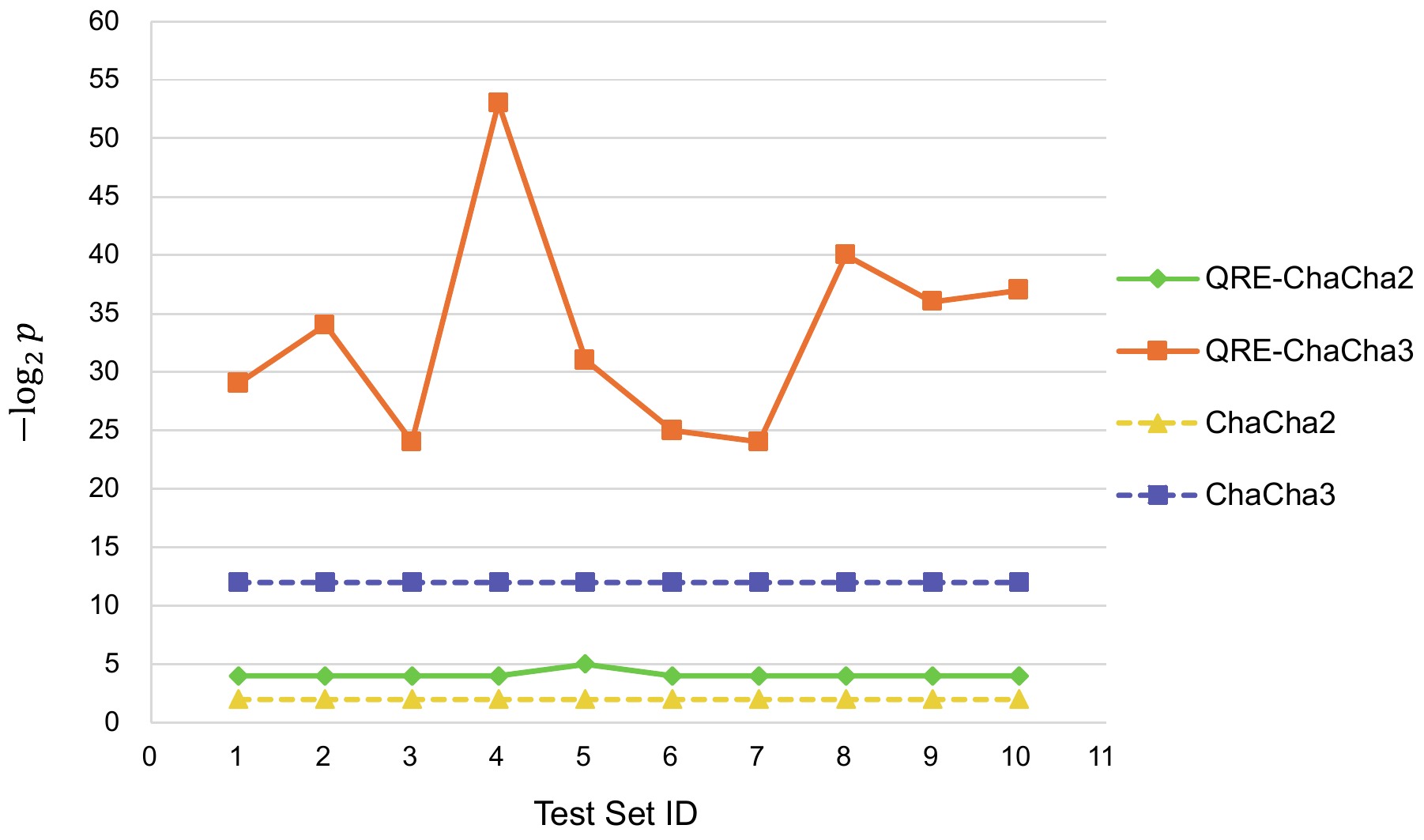}
	\caption{Differential probabilities of 2-round and 3-round QRE-ChaCha (10 Sets)}\label{fig4}
\end{figure}

In order to analyze the specific contribution of quantum random numbers to the security of QRE-ChaCha, we selected 10 independently generated pairs of quantum random numbers and computed the differential values between each pair. These differentials were used as fixed constraints in the automated differential search process. Based on this setup, we determined the optimal differential trails and corresponding differential probabilities after 2 and 3 rounds of iteration. The upper bounds of the differential probabilities are illustrated in Figure \ref{fig4}. The upper bound of the 2-round differential probability remains stable at approximately $2^{-4}$, while the 3-round differential probability fluctuates between $2^{-24}$ and $2^{-53}$. Furthermore, we computed the average over the 10 sets to obtain representative results, with the 2-round average upper bound at approximately $2^{-4}$ and the 3-round counterpart at around $2^{-25}$.

Based on the final search results and subsequent averaging, as shown in Table \ref{tab2}, the 2-round average upper bound on the differential trail probability for the QRE-ChaCha is approximately $2^{-4}$, while that of the 3-round case is approximately $2^{-25}$. Therefore, the number of effective differential trails (with $Pr > 2^{-512}$) does not exceed $3{\times}20+2{\times}3=66$ rounds. The upper bound on the 20-round differential trail probability is approximately $2^{-154}$. Under the same testing conditions and methodology, the 2-round differential trail probability upper bound for the original ChaCha is $2^{-2}$, and for 3 rounds, it is $2^{-12}$. Accordingly, the number of effective differential trails ($Pr > 2^{-512}$) for ChaCha does not exceed $3{\times}42+2{\times}4=134$ rounds, and the upper bound on the 20-round probability is $2^{-74}$. Hence, under the analysis framework adopted in this work, QRE-ChaCha demonstrates significantly stronger resistance against differential cryptanalysis compared to the original ChaCha.

\begin{table}[htb]
    \centering
    \caption{Average Differential Probabilities of QRE-ChaCha and ChaCha}\label{tab2}
    \begin{tabular}{|c|c|c|}
        \hline
        \textbf{Algorithm} & \textbf{Rounds} & \boldmath$\log_2{p}$ \\
        \hline
        \multirow{2}{*}{QRE-ChaCha}
        & 2 & $-4$ \\
        & 3 & $-25$ \\
        \hline
        \multirow{2}{*}{ChaCha}
        & 2 & $-2$ \\
        & 3 & $-12$ \\
        \hline
    \end{tabular}
\end{table}

\section{Statistical Randomness Testing}\label{sec6}

To further validate the quality of the QRE-ChaCha's output, we comprehensively evaluate the randomness characteristics of the QRE-ChaCha using the NIST Statistical Test Suite and the Chinese Cryptographic Randomness Test Standard in this section. Specifically, the tests are performed using the NIST SP 800-22 Rev. 1 \cite{ref45}, and the open-source randomness toolkit project\cite{ref46}.

The NIST Statistical Test Suite is a widely adopted standard consisting of 15 tests designed to assess the randomness of binary sequences generated by hardware- or software-based cryptographic random or pseudorandom number generators. The Chinese cryptographic randomness test conforms to the national standard GM/T 0005-2021: Specification for Randomness Testing\cite{ref47}, which also includes 15 tests. Among them, 11 are consistent with those in the NIST suite, including the frequency test, block frequency test, runs test, longest run of ones in a block test, binary matrix rank test, discrete Fourier transform test, Maurer's universal statistical test, linear complexity test, overlapping template matching test, approximate entropy test, and cumulative sums test. Although the same tests are used, the two may differ slightly in implementation details. In addition, the Chinese standard includes four specialized test items: the poker test, run distribution test, binary derivation test, and autocorrelation test. Together, these two test suites ensure that the generated random number sequences exhibit strong statistical randomness, thereby meeting the requirements of cryptographic and other randomness-dependent applications. The test results demonstrate that the QRE-ChaCha introduces no flaws in the randomness of its keystream. The overall testing procedure adopted in this work is summarized as follows:

\begin{itemize}
    \item This test is based on the 8-round version of the QRE-ChaCha, which is the minimum number of rounds permitted under our evaluation criteria. Using randomly generated seed keys, we encrypt identical plaintexts with QRE-ChaCha8 to produce 10{,}000 keystream sequences, each with a length of 1{,}000{,}000 bits.
    \item These 10{,}000 keystream sequences are subjected to both the NIST Statistical Test Suite and the Chinese National Cryptographic Randomness Test Suite. During testing, all parameters recommended by NIST and the Chinese standard are adopted. The significance level for both tests is set to $0.01$, and the uniformity significance level is set to $0.0001$. The results are analyzed to determine whether the keystreams generated by QRE-ChaCha exhibit statistically strong randomness. The test results from both suites are summarized in Table \ref{tab3} and Table \ref{tab4}.
\end{itemize}

\begin{table}[htbp]
    \centering
    \caption{NIST randomness test results for 1000 sets of QRE-ChaCha8 keystreams}\label{tab3}
    \begin{tabular}{|c|c|c|}
        \hline
        NIST Test Item & Pass Count & P-Value \\
        \hline
        Frequency                & 982 & 0.187581 \\
        Block Frequency          & 988 & 0.751866 \\
        Cumulative Sums          & 983 & 0.435430 \\
        Runs                     & 994 & 0.062821 \\
        Longest Run of Ones      & 984 & 0.747898 \\
        Rank                     & 990 & 0.784927 \\
        FFT                      & 986 & 0.803720 \\
        Non-overlapping Template & 982 & 0.940080 \\
        Overlapping Template     & 990 & 0.117432 \\
        Universal Statistical    & 990 & 0.012829 \\
        Approximate Entropy      & 990 & 0.345650 \\
        Serial                   & 992 & 0.899171 \\
        Linear Complexity        & 987 & 0.115387 \\
        \hline
    \end{tabular}
\end{table}

\begin{table}[htbp]
    \centering
    \caption{Randomness test results of 10{,}000 sets of QRE-ChaCha8 keystreams using GM/T 0005-2021}\label{tab4}
    \begin{tabular}{|c|c|c|}
        \hline
        GM/T 0005-2021 Test Item & Pass Count & P-Value \\
        \hline
        Single Bit Frequency                 & 9884 & 0.862398 \\
        Block Frequency ($m=10000$)          & 9902 & 0.969009 \\
        Poker Test ($m=4$)                   & 9889 & 0.469806 \\
        Poker Test ($m=8$)                   & 9910 & 0.362434 \\
        Overlapping Template ($m=3$, P1)     & 9890 & 0.978538 \\
        Overlapping Template ($m=3$, P2)     & 9901 & 0.211848 \\
        Overlapping Template ($m=5$, P1)     & 9918 & 0.610070 \\
        Overlapping Template ($m=5$, P2)     & 9915 & 0.906880 \\
        Total Runs                           & 9915 & 0.113239 \\
        Run Distribution                     & 9900 & 0.399442 \\
        Max Run of 1s ($m=10000$)            & 9900 & 0.386748 \\
        Max Run of 0s ($m=10000$)            & 9902 & 0.650860 \\
        Binary Derivation ($k=3$)            & 9905 & 0.699313 \\
        Binary Derivation ($k=7$)            & 9889 & 0.669151 \\
        Autocorrelation ($d=1$)              & 9915 & 0.073281 \\
        Autocorrelation ($d=2$)              & 9898 & 0.187378 \\
        Autocorrelation ($d=8$)              & 9902 & 0.128354 \\
        Autocorrelation ($d=16$)             & 9893 & 0.846168 \\
        Matrix Rank                          & 9902 & 0.008056 \\
        Cumulative Sums (Forward)            & 9885 & 0.394370 \\
        Cumulative Sums (Backward)           & 9889 & 0.447116 \\
        Approximate Entropy ($m=2$)          & 9890 & 0.981469 \\
        Approximate Entropy ($m=5$)          & 9915 & 0.216485 \\
        Linear Complexity ($m=500$)          & 9882 & 0.526907 \\
        Linear Complexity ($m=1000$)         & 9887 & 0.155238 \\
        Maurer Universal ($L=7$, $Q=1280$)   & 9892 & 0.621922 \\
        Discrete Fourier Transform ($m=500$) & 9892 & 0.294959 \\
        \hline
    \end{tabular}
\end{table}

To constrain the file size during testing, only 1{,}000 keystream sequences, each with a length of 1{,}000{,}000 bits, were used for the NIST randomness evaluation. Table \ref{tab3} presents selected results. For the NonOverlappingTemplate test, only the result corresponding to the template with the minimum number of samples passing the significance level is shown. Results from the RandomExcursions and RandomExcursionsVariant tests are omitted due to the large number of sub-tests, but the keystreams successfully passed all items within these two test categories.

According to the test results, the keystreams generated by QRE-ChaCha8 successfully passed both the NIST and the GM/T 0005-2021 randomness test suites. This indicates that the keystreams exhibit strong randomness properties and meet the requirements for cryptographic applications. Furthermore, the QRE-ChaCha explicitly considers randomness enhancement during its design. By optimizing the original ChaCha structure and incorporating quantum random numbers, additional confusion is introduced, thereby significantly improving the statistical quality of the generated keystreams. The output sequences exhibit excellent distribution uniformity, with no observable statistical patterns exploitable by adversaries. This effectively strengthens resistance against statistical and cryptanalytic attacks, and further demonstrates that the integration of quantum random numbers contributes to enhancing the overall security of the cipher.

\section{Performance Evaluation}\label{sec7}

In this section, to quantify the computational efficiency of the QRE-ChaCha, we measure its performance by timing the encryption of fixed-size files. The testing environment is configured as follows: an AMD Ryzen 7 5700U processor with Radeon Graphics, clocked at 1.80 GHz, running a 64-bit Windows 10 Enterprise Edition (version 22H2), with 16 GB of RAM in an x64 architecture. The encryption and decryption operations were implemented at the software level using the C programming language.

In the performance evaluation, we used the 8-round versions of both QRE-ChaCha and ChaCha for encryption and decryption comparison tests, with the 20-round version of ChaCha included as a reference. For each cipher, randomly generated seed keys were used to encrypt the same files of sizes 10 MB, 20 MB, 30 MB, 40 MB, and 50 MB. Each file size was tested 5 times, and the average encryption time was taken as the final performance metric. The results are summarized in Table \ref{tab5}. Based on the performance test results, the encryption time of QRE-ChaCha8 is almost identical to that of ChaCha8.

\begin{table}[htbp]
    \centering
    \caption{Encryption time comparison of QRE-ChaCha8, ChaCha8, and ChaCha20}\label{tab5}
    \begin{tabular}{|c|c|c|c|}
        \hline
        \multirow{2}{*}{File Size} & \multicolumn{3}{c|}{Encryption Time (s)} \\
        \cline{2-4}
        & QRE-ChaCha8 & ChaCha8 & ChaCha20 \\
        \hline
        10 MB & 0.1037854 & 0.1051830 & 0.2025330 \\
        20 MB & 0.2096104 & 0.2115156 & 0.4061916 \\
        30 MB & 0.3118018 & 0.3147998 & 0.6116970 \\
        40 MB & 0.4168038 & 0.4228406 & 0.8162400 \\
        50 MB & 0.5273160 & 0.5308238 & 1.0211580 \\
        \hline
    \end{tabular}
\end{table}

It is worth noting that the performance evaluation presented in this work does not account for the time required to generate quantum random numbers. This is because modern QRNGs are capable of delivering secure random numbers at rates exceeding 20 Gbps\cite{ref48,ref49}, which can be pre-stored in the memory module, rendering their impact on overall encryption time negligible. Therefore, the performance tests conducted in this work focus solely on the algorithmic structure. The results demonstrate that the integration of quantum randomness does not degrade the encryption or decryption efficiency of QRE-ChaCha. On the contrary, the algorithm retains the high performance of ChaCha while achieving an improvement in security.

\section{Conclusion}\label{sec8}

As a conclusion, we propose an innovative stream cipher, QRE-ChaCha, which enhances the security of the classical ChaCha cipher by introducing quantum random numbers. The core enhancement is achieved through a two-stage process. Firstly, quantum random numbers are XORed into the initial constant block of the state matrix to bolster the randomness of the seed key. Secondly, during each odd-numbered round, the first 128 bits of the intermediate state are further strengthened with additional quantum random numbers, also via XOR. These periodic injections, combined with the strong diffusion properties of the round function, ensure that the true randomness from the quantum source permeates the entire state matrix, thereby improving the cipher's overall cryptographic security.

In terms of security analysis, this work first examines the theoretical security advantages of QRE-ChaCha by analyzing the physical principles of quantum random number generation and attack resilience. Subsequently, differential cryptanalysis was performed using a SAT-based automated search approach. Compared with the original ChaCha cipher, QRE-ChaCha demonstrates significantly enhanced resistance to differential attacks. Specifically, the upper bounds of the average differential trail probabilities for 2 and 3 rounds of QRE-ChaCha are $2^{-4}$ and $2^{-25}$, respectively, whereas those of the original ChaCha are $2^{-2}$ and $2^{-12}$. For 20 rounds, QRE-ChaCha achieves an upper bound of $2^{-154}$, considerably lower than the $2^{-74}$ bound observed in ChaCha.

Regarding keystream randomness, this work evaluated QRE-ChaCha using both the NIST Statistical Test Suite and the GM/T 0005-2021 standard for randomness testing. The results confirm that the keystreams generated by QRE-ChaCha exhibit strong statistical randomness and meet the requirements for cryptographic applications.

For performance evaluation, this work measures the encryption speed of QRE-ChaCha on input files of varying sizes to assess its overall encryption and decryption efficiency. Experimental results indicate that QRE-ChaCha maintains high performance. Specifically, the encryption time of QRE-ChaCha8 is nearly identical to that of ChaCha8, suggesting that the integration of quantum random numbers introduces no noticeable performance overhead. This observation holds consistently across encryption tasks involving files ranging from 10MB to 50MB in size.

In summary, the proposed QRE-ChaCha cipher preserves the high performance of the original ChaCha cipher while significantly enhancing its security through the integration of quantum random numbers. The improvements are evident in both the cipher's resistance to differential attacks and the high randomness quality of its keystream. Performance evaluations further confirm the cipher's practical viability. Moreover, QRE-ChaCha can also be regarded as a quantum randomness expansion scheme, offering new prospects for broader applications of quantum-generated randomness in cryptographic and computational contexts.


\end{document}